\documentclass[11pt]{article}

\usepackage[dvipsnames]{xcolor}

\usepackage{times}

\usepackage{dsfont}
\usepackage{amsmath}
\usepackage{amsthm}
\usepackage{amssymb}
\usepackage{mathtools}

\usepackage{thmtools} 
\usepackage{thm-restate}

\usepackage{graphicx}
\usepackage{url}
\usepackage{nicefrac}

\usepackage{enumitem}
\setlist[enumerate]{nosep}

\usepackage{subcaption}
\usepackage{xspace}
\usepackage[compact]{titlesec}
\usepackage[normalem]{ulem}
\usepackage{comment}

\usepackage{algorithm}
\usepackage[noend]{algpseudocode}

\usepackage{tikz} 
\usetikzlibrary{arrows.meta, positioning, calc}

\usepackage{pgf}
\usetikzlibrary{patterns}
\usetikzlibrary{backgrounds}
\usetikzlibrary{scopes}
\usetikzlibrary{arrows, automata, positioning}
\usetikzlibrary{decorations.markings}
\usepackage[T1]{fontenc}
\usepackage{calc}
\usepackage{color}
\usepackage{lipsum}

\usepackage[bottom=1.5in,top=1in,left=1in,right=1in]{geometry}
\makeatother
\usepackage{color}

\definecolor{blueblack}{rgb}{0,0,.7}

\newcounter{sideremark}

\definecolor{Darkblue}{rgb}{0,0,0.4}
\definecolor{Brown}{cmyk}{0,0.61,1.,0.60}
\definecolor{Purple}{cmyk}{0.45,0.86,0,0}
\definecolor{brickred}{rgb}{0.8, 0.25, 0.33}
\usepackage[colorlinks,citecolor=blue,linkcolor=red,bookmarks=true]{hyperref}
\usepackage{cleveref}

\theoremstyle{plain}
\newtheorem{theorem}{Theorem}
\newtheorem{proposition}[theorem]{Proposition}
\newtheorem{lemma}[theorem]{Lemma}
\newtheorem{corollary}[theorem]{Corollary}
\newtheorem{claim}[theorem]{Claim}

\theoremstyle{definition}

\theoremstyle{remark}

\newtheorem*{remark*}{Remark}

\newcommand{\NN}{{\mathbb N}}
\newcommand{\EE}{{\mathbb E}}

\newcommand{\pr}{\mathbb{P}}

\newcommand{\cala}{\mathcal{A}}
\newcommand{\calb}{\mathcal{B}}
\newcommand{\calc}{\mathcal{C}}

\newcommand{\calx}{\mathcal{X}}

\newcommand{\OPT}{\mathsf{OPT}}
\newcommand{\opt}{\mathsf{opt}}

\newcommand{\density}{\mathsf{Den}}
\newcommand{\DkS}{D$k$S\xspace}
\newcommand{\DVP}{DVP\xspace}

\newcommand{\tclose}{\mathcal{C}}
\newcommand{\treduc}{\mathcal{R}}

\newcommand{\indic}{\boldsymbol{1}}

\DeclareMathOperator*{\argmax}{arg\;max}

\DeclareMathOperator{\polylog}{polylog}

\usepackage[colorinlistoftodos,textsize=tiny,textwidth=2cm,color=green!50!gray]{todonotes} 

\ifdefined\DEBUG
  \newcommand{\ant}[1]{\textcolor{blue}{#1}}
  \newcommand{\wes}[1]{\textcolor{ForestGreen}{#1}}

\newcommand{\anote}[1]{\todo[color=blue!100!black!50]{A: #1}}
\newcommand{\wnote}[1]{\todo[color=green!100!black!50]{W: #1}}

\else
  \newcommand{\ant}[1]{#1}
  \newcommand{\wes}[1]{#1}

  \newcommand{\anote}[1]{}
  \newcommand{\wnote}[1]{}
\fi

\author{Antoine Tinguely
\\
{\small\texttt{antoine.tinguely@tum.de}}
\and
G. Wessel van der Heijden
\\
{\small\texttt{wessel.heijden@tum.de}}}

\title{
Approximation and Hardness for the Data Validation Problem
}

\date{}
\begin{document}

\maketitle

\begin{center}
    {\small Technical University of Munich, Germany}
    \vspace{10pt}
\end{center}

\begin{abstract}
    In scheduling, precedence constraints stipulate that certain tasks can only be scheduled after the completion of other tasks, and have been studied extensively in theory and applications.
    In many practical cases, some presumed precedences are redundant, so it can be beneficial to invest some resources to challenge the validity of such constraints. If a constraint turns out to be unnecessary, discarding it can drastically improve the optimum of the scheduling instance.
    
    To model such scenarios, Finnah, G{\"o}nsch and Otto [Eur.\ J.\ Oper.\ Res.\ 2025] recently introduced the Data Validation Problem (DVP):
    Given a transitive directed acyclic graph $G$ on $n$ vertices (resp.\ tasks) and $m$ edges (resp.\ precedences) and a probing budget, a policy can successively pay a probing fee and probe an edge in $G$ as long as the total probing fees do not exceed the budget.
    If a probe succeeds, the targeted edge $e$ is removed from $G$ and we receive a reward; if the probe fails, the edge stays in $G$ and cannot be probed again.
    We seek a policy with maximum expected cumulative reward.
    Crucially, $G$ must stay transitive throughout the probing process, so a policy can only challenge the precedence between consecutive tasks, that is, only probe edges in the transitive reduction of the current precedence graph.

    While DVP has been explored using heuristic methods, little is known about the worst-case complexity and approximability of the problem.
    We initiate the theoretical study of DVP by proving that DVP is strongly NP-hard and W[1]-hard with the probing budget $T$ as a parameter, even in the deterministic case (when every probe has probability $0$ or $1$ to succeed) and with unit probing costs and rewards.
    For the same setting, we give an approximation-preserving reduction from the Densest $k$-Subgraph (D$k$S) Problem, with the consequence that there is no polynomial-time $2^{(\log n)^{1-\delta}}$-approximation algorithm and no FPT $\sqrt{T}$-approximation algorithm for DVP, assuming the Almost-Exponential Time Hypothesis and the Strongish Planted Clique Hypothesis, respectively.

    We also present the first known policies with provable worst-case guarantees for DVP, namely an $m$-approximation algorithm for instances with uniform probabilities and probing costs, a $2T/p_{\min}^T$-approximation algorithm for general instances, where $p_{\min}$ is the minimum probing success probability, and an exact FPT algorithm on tournament graphs, in the deterministic setting.
\end{abstract}


\thispagestyle{empty}
\newpage
\setcounter{page}{1}

\section{Introduction}\label{sec:introduction}

Scheduling under precedence constraints is a classic topic in theoretical computer science and operations research. 
In a precedence-constrained scheduling instance, a set of tasks is given together with a partial order $\prec$, which is the set of \emph{precedence constraints}.
\ant{The partial order can also be encoded as a so-called precedence graph.}
Given two tasks $t,t'$, whenever $t\prec t'$, task $t$ must be completed before task $t'$ can start. 
Tasks that are incomparable in the partial order $\prec$ may be processed in any order. 
The focus of the literature for scheduling with precedence constraints has considered the partial order as an input and has been primarily focused on scheduling tasks according to some objective function depending on when tasks are assigned on a time line. 
Many variations of precedence constraints have been studied, where the structure of the partial order $\prec$ can strongly affect both the computational complexity of the problem and the optimal objective value \cite{prot2018survey}. 
A more basic question has received much less attention: how is the precedence structure obtained in the first place?
In practice, the precedence constraints are rarely given explicitly with full confidence and must first be determined. 
This is a question that has been considered in the operations research community as the \emph{Data Validation Problem (DVP)} introduced by Finnah, G{\"o}nsch and Otto~\cite{finnah2025overcoming}. 
However, to the best of our knowledge, the computational complexity or approximation bounds for this problem have not been studied from a theoretical perspective. 
In this paper, we initiate the theoretical study of the problem of determining the precedence structure of a scheduling problem. 
We give complexity results and approximation guarantees for this problem and several natural variants.

We formally describe \ant{DVP following the definition in \cite{finnah2025overcoming}}. 
We are given a \emph{probing budget} $T \in \NN$ and a directed acyclic graph (DAG) $G=(V,E)$ which is also transitive, that is, $(u,v)\in E$ and $(v,w)\in E$ imply $(u,w)\in E$.
\ant{Let $n\coloneqq |V|$ and $m\coloneqq |E|$.}
Each edge $e\in E$ has a \emph{weight} $w_e\in\NN_0$, a \emph{probing cost} $\tau_e\in \NN_0$, and a \emph{success probability} $p_e\in[0,1]$, to which is tied a random variable $\Omega_e \sim \text{Bernoulli}(p_e)$. 
We assume the random variables $\{\Omega_e\}_{e\in E}$ to be (mutually) independent.

Iteratively, an (adaptive) \emph{probing policy} can probe an edge $e\in E$ by paying $\tau_e$, i.e., by decreasing the probing budget by $\tau_e$.
Then, we reveal the outcome of $\Omega_e$ (which is unknown before the probe). If the probe \emph{succeeds}, i.e., in the event that $\Omega_e =1$, we collect $w_e$ as a reward, and $e$ is removed from $E$; otherwise, if the probe \emph{fails}, i.e., if $\Omega_e=0$, we collect no reward, $e$ stays in $E$ and cannot be probed anymore.
Crucially, we impose that an edge can only be probed if it lies in the \emph{transitive reduction of $E$}, which is the set of all edges $(u,v) \in E$ such that $E\setminus\{(u,v)\}$ contains no (directed) $u$-$v$-path,%
\footnote{When we speak of paths on a DAG, we always mean \emph{directed} paths.}
which we denote by $\treduc(E)$. The transitive reduction of $G$ can be computed in time $O(n^{2.4})$ \cite{aho1972transitive,bang2008digraphs} \ant{statically and be updated in time $O(n^{1.528} + m)$ at each edge deletion \cite{goranci2025fully}}.
The policy must halt if every edge in $\treduc(E)$ has already been probed or if $\tau_e>T$ for any unprobed edge $e\in \treduc(E)$.
Thereby, a policy $\pi$ yields the expected total reward $W(\pi) \coloneqq \EE\big[\sum_{e\in E} \indic_{\{\pi \text{ successfully probes }e\}}\big]$, where $\indic_C$ denotes the indicator function of an event $C$; the goal is to find a policy $\pi^*$ with maximum expected reward, i.e., $\pi^* \in \argmax_{\text{policy }\pi} W(\pi)$.

\ant{We remark that we allow policies to sample random variables other than $\{\Omega_e\}_{e\in E}$ but that are independent of $\{\Omega_e\}_{e\in E}$.}
\ant{For any $E' \subseteq E$, we write $\tau(E') \coloneqq \sum_{e\in E'}\tau_e$, $w(E') \coloneqq \sum_{e\in E'}w_e$ and $p(E') \coloneqq \prod_{e\in E'}p_e$.}

\ant{Intuitively, the assumption of only probing edges in the transitive reduction is motivated by the interpretation of $G$ as a precedence graph.
If a precedence graph $G$ contains a path $P=(u_1, u_2, \dots, u_k)$, by transitivity, it also contains the edge $e=(u_1,u_k)$, so removing $e$ but no edge in $P$ cancels transitivity, meaning that this framework no longer models the removal of redundant precedences in a meaningful way.}

Determining the precedence constraints between tasks is a problem that is often faced in practice. 
The problem arises in many industries with business processes subject to hard constraints such as precedence constraints, including product design, assembly and manufacturing, project scheduling, construction, and logistics \cite{finnah2025overcoming,wang2009assembly}. 
In such processes, the relation between every pair of tasks is rarely documented explicitly and \ant{an
expert must be consulted to determine it.} 
Moreover, Otto and Otto \cite{otto2014multiple} showed that probing direct precedence relations can increase the probability of being able to remove the edge in automotive assembly processes. 
Wang et al.\ \cite{wang2009assembly} investigated the Assembly Sequence Problem, a problem similar to \DVP{}, which is a well-studied problem in the operations research community. 
Finnah et al.\ \cite{finnah2025overcoming} recently expanded on the model by imposing constraints on the available expert time or cost.
They showed that solving the problem, even approximately, can significantly improve business outcomes. 
Other questions regarding the complexity or approximation guarantees remained open.

\subsection{Our results}\label{sec:results}

We significantly strengthen the hardness result in \cite{finnah2025overcoming}. Via an FPT reduction from the $k$-Clique problem, we show that \DVP is strongly NP-hard and W[1]-hard, even in the deterministic setting (when all probabilities are either 0 or 1) and with unit costs and weights.

\begin{theorem}\label{thm:k-clique}
    \DVP is strongly NP-hard and W[1]-hard when parametrized in $T$, even if $\tau_{e} = p_e = 1$ and $w_{e} \in \{0,1\}$ for every edge $e \in E$.
\end{theorem}

We also achieve an approximation-preserving reduction to the Densest $k$-Subgraph Problem (\DkS). Thereby, using known (parametrized) inapproximability results for \DkS, we achieve strong bounds for \DVP as well. 
\ant{The bounds assume the Almost-Exponential Time Hypothesis (Almost-ETH)~\cite{manurangsi2026almost} (a slightly weaker assumption than the Exponential Time Hypothesis) and the Strongish Planted Clique Hypothesis (SPCH)~\cite{manurangsi2021strongish}, respectively.}

\begin{theorem}\label{thm:dks}
    If there is an $\alpha$-approximation for \DVP{} ($\alpha$ can depend on $T$, $m$ or $n$), even restricted to the special case that $\tau_{e} = p_e = 1$ and $w_{e} \in \{0,1\}$ for every edge $e \in E$, there is an $\alpha \cdot \big(4+O(1/k)\big)$-approximation for \DkS{} (with $T = \Theta(\opt)$, where $\opt$ is the optimum of the \DkS{} instance).
    As a consequence, even in this special case,
    \begin{enumerate}[label={(\arabic*)}]
        \item\ant{\label{item:poly-almost-eth} assuming the Almost-ETH, for every $\delta>0$ there is no polynomial-time $2^{(\log n)^{1-\delta}}$-approximation algorithm for \DVP;
        }
        \item\label{item:fpt-spch} assuming the SPCH, there is no $o\big(\sqrt{T}\big)$-approximation algorithm for \DVP in time $g(T) \cdot n^{O(1)}$ for any function $g$.
    \end{enumerate}
\end{theorem}


As pointed out in \cite{finnah2025overcoming}, a greedy policy that only considers the currently probable edges, e.g., that always picks as a next probe the best cost-per-value edge $\argmax_{e\in \treduc(E)} \frac{w_e}{\tau_e}$, performs arbitrarily badly, even for very simple instances.

In contrast, we present an adaptive policy which is an $m$-approximation in the case of uniform costs and success probabilities and arbitrary weights and arbitrary DAGs.

\begin{theorem}\label{thm:adaptive}
    For any fixed $p \in (0,1]$, there exists an $m$-approximation algorithm for DVP with running time $O(mn\cdot \min\{m,T\})$ in the case $\tau_e =1$, $p_e = p$ for every $e\in E$.
\end{theorem}

We also present a more general policy for arbitrary instances of DVP with an approximation factor that depends on the budget $T$ and the minimum non-zero success probability.

\begin{theorem}\label{thm:greedy}
	There exists an $\frac{2T}{p_{\min}^T}$-approximation algorithm for DVP with running time $O(m^2n)$, where $p_{\min} = \min \{p_e:\; e\in E, \; p_e\neq 0\}$.
    \anote{can maybe assume $p_e\neq 0$ by setting $\tau_e > T$ whenever necessary}
\end{theorem}

Notice that in the deterministic case, the latter algorithm is a $2T$-approximation, which comes close to the lower bound in Theorem~\ref{thm:dks}.
\ant{We remark that none of the algorithms above actually requires $G$ to be transitive.}

For the special case of \emph{total order}, that is, if $G$ forms a (acyclic) \emph{tournament} (for every $u,v \in V$ either $(u,v) \in E$ or $(v,u)\in E$), or equivalently, if $\treduc(E)$ is a path on $V$, in the deterministic setting, the optimal policy is computable in FPT time.

\begin{theorem}\label{thm:path-case}
    If $G$ is a (acyclic) tournament, assuming $p_e =1$ and $\tau_e > 0$ for every $e\in E$, the probing set with maximum weight can be computed in time $2^{O(T\log T)} \cdot n^{O(1)}$.
\end{theorem}

\subsection{Related work}
In the Pandora's Box problem (PBP), as in \DVP{}, each item has an associated probing cost and a random reward.
However, the objective is to maximize the expected maximum sampled reward minus the total probing cost.
An optimal policy can be found in polynomial time for tree-like precedences \cite{boodaghians2020pandora,weitzman1979pandora}, however, it is NP-hard to approximate PBP above a factor $0.9997$ for general precedences or matroid constraints.

PBP belongs to the family utility/disutility-optimization problems, for which there are generic greedy constant-factor approximation algorithms \cite{singla2018price}.
Similarly, if the objective function of a stochastic problem is adaptive submodular (the probabilistic equivalent of submodular functions), which is unfortunately not the case for \DVP{}, there are simple greedy policies that are constant-factor approximations \cite{golovin2011adaptive}.

Another problem similar to \DVP{} is the Stochastic Probing problem \cite{gupta2013stochastic}.
Here, as in \DVP{}, every element to be probed has a weight and a success probability, and the goal is to maximize the expected total weight of the successively probed elements.
The rule is that during the entire probing process, the set of probed and successfully probed elements, respectively, must satisfy packing constraints like Knapsack or matroid constraints, and the approximation ratio of the policy in \cite{gupta2013stochastic} depends on the complexity of the constraints.
For example, if the sets have to belong to the intersection of $k$ matroids, the approximation ratio is $\Omega(1/k)$.

The $s$-$t$-unreliability problem asks for the probability that a pair of vertices u and v become disconnected when each edge $e$ of a graph $G$ is independently removed with probability $p_e$.
By Proposition~\ref{prop:FUS-cut}, it is the same as to ask, assuming $T=\infty$, whether the edge $(u,v)$ can successively be probed or not, so the problem can be helpful to provide success bounds for \DVP{} algorithms.
Unfortunately, the problem is \#BIS-hard \cite{feng2024FPRAS}, meaning that there is most likely no (randomized) PTAS for the problem \cite{dyer2004relative}.

\section{Hardness results}

We construct a parametrized and approximation-preserving reduction from $k$-Clique and \DkS, respectively.
Both problems are strongly NP-hard and W[1]-hard \cite{cygan2015parameterized,garey1979computers,lanciano2024survey}.

 We recall that in \DkS, the goal is, given an undirected graph \wes{$H=(U,F)$}, to find the induced subgraph of \wes{$H$} with $k$ vertices and with a maximum number of edges, that is, to find $\argmax_{S \subseteq U, |S|=k}\density(S)$, where $\density(S) \coloneqq \ant{F}[S]|/\binom{|S|}{2}$ is the \emph{density of $S$} ($F[S]$ is the edge set of the subgraph induced by $S$).
The $k$-Clique problem is to check whether the optimal solution to \DkS is a clique.
\ant{For a positive integer $m$, we denote $[m] \coloneqq \{1, 2, ..., m\}$. For a set $X$, $|X|$ denotes its cardinality.}

Let $k\in \NN$ and consider an undirected graph $H = (U,F)$.
Let $D\in \NN$ to be fixed later.
We consider the \DVP instance $G_H=(V_H,E_H)$ with vertex and edge sets
\begin{equation*}
    V_H = \{s_{e}\}_{e\in F} \cup \{t^i_u\}_{u\in U}^{i \in [D]} \cup \{z\}
    \quad\text{and}\quad
    E_H = \{p_{e,u}^i\}_{e\in F, e\in u}^{i\in [D]} \cup \{q_u^i\}_{u\in U}^{i\in [D]} \cup \{r_e\}_{e\in F},
\end{equation*}
where edges are defined as $p^i_{e,u} = (s_e, t^i_u)$ (where $u$ is incident to $e$), $q_u^i = (t_u^i,z)$ and $r_{e} = (s_{e},z)$.
All edges in $E_H$ have success probability and probing cost $1$, the edges $r_e$ have weight $1$, other edges in $E_H$ have weight $0$.
Notice that only the edges $r_{e}$ can yield profit. To probe $r_{e}$, where $e = \{u,v\}\in F$, it is \ant{necessary and sufficient to probe $q^i_w$ or $p_{e,w}^i$ for every $i\in [D]$ and every $w\in\{u,v\}$.}
Clearly, $G_H$ is a transitive DAG.

\begin{proof}[Proof of Theorem~\ref{thm:k-clique}]
    Set $D=1$ and $T = k+ \binom{k}{2}$. We show that the construction above is a parametrized reduction (see \cite{cygan2015parameterized} for the definition) from $k$-Clique to \DVP.
    More precisely, we show that the \wes{graph} $\ant{H}$ has a $k$-clique if and only if $w(\OPT) = \binom{k}{2}$, where $\OPT$ is the optimal solution \ant{(the edges in $E_H$ removed by an optimal policy)} of the \DVP instance $G_H$.
    \ant{Let $S \coloneqq \{u\in U:\; p^1_{e,u} \in \OPT \text{ for some }e\in F\}$. Clearly, $\OPT' \coloneqq (\OPT\setminus \{p^1_{e,u}\}_{u\in S, e\in u}) \cup \{q^1_u\}_{u\in S}$ is feasible and also optimal since $w(\OPT') = w(\OPT) = |\OPT\cap \{\ant{r}_{e}\}_{e\in F}|$ and $|\OPT'| \leq |\OPT|$.}

    Assume $w(\OPT') \geq \binom{k}{2}$, and so $\ell \coloneqq |\OPT' \cap \{q^1_u\}_{u\in U}| \leq  T - \binom{k}{2} = k$.
    For each $u \in U$, let $N_u \coloneqq \{r_{e} \in \OPT:\; u\in e\}$. We have
    \begin{equation*}
        k(k-1)
        \leq 2w(\ant{\OPT'})
        = \sum_{u\in U} |N_u|
        \leq \ell(\ell-1)
        \leq k(k-1).
    \end{equation*}
    The second inequality holds because there are at most $\ell$ many vertices $u$ with $N_u \neq \emptyset$ and $|N_u \cap N_v| \leq 1$ for every pair $u, v \in U$.
    All inequalities are tight, so $w(\ant{\OPT'}) = \binom{k}{2}$, and we have exactly $k$ many vertices $u\in U$ with $N_u \neq \emptyset$ and $|N_u| = k-1$. Therefore, those vertices form a $k$-clique in \wes{$H$}.
    
    Conversely, if \wes{$H$} has a $k$-clique $S \subseteq F$, then probing the edges $\{s_{e}\}_{e\in F[S]} \cup \{q^1_u\}_{u\in U}$ needs exactly $T$ probes and yields a profit $\binom{|S|}{2} = \binom{k}{2}$, thus $w(\ant{\OPT'}) \geq \binom{k}{2}$, and therefore $w(\ant{\OPT'}) = \binom{k}{2}$ as shown above.
\end{proof}

\begin{figure}[t]
    \centering

\begin{tikzpicture}[
  >=Stealth,
  vertex/.style    ={circle, draw, fill=blue!10,   minimum size=7mm, inner sep=0pt, font=\small},
  source/.style    ={circle, draw, fill=orange!15, minimum size=7mm, inner sep=0pt, font=\small},
  sink/.style      ={circle, draw, fill=gray!25,   minimum size=8mm, inner sep=0pt, font=\small},
  vertexU/.style   ={circle, draw, fill=gray!10,   minimum size=7mm, inner sep=0pt, font=\small, draw=gray},
  lit/.style       ={-{Stealth}, thick, green!45!black},
  litU/.style      ={-{Stealth}, thick, gray},
  dom/.style       ={-{Stealth[length=2mm]}, thin, gray!85},
  prize/.style     ={-{Stealth}, dashed, thick, red!70!black},
  prizeU/.style    ={-{Stealth}, dashed, thick, gray},
]

\begin{scope}[xshift=0cm]

\node[vertex] (a) at (-0.8,3.6)  {$a$};
\node[vertex] (c) at (0.8,3.6)   {$c$};
\node[vertex] (b) at (0,2.4)     {$b$};
\node[vertex] (d) at (0,0.8)     {$d$};

\draw[dashed, thick, red!70!black] (a)--(c);
\draw[dashed, thick, red!70!black] (a)--(b);
\draw[dashed, thick, red!70!black] (c)--(b);
\draw[thick, gray] (b)--(d);

\end{scope}

\begin{scope}[xshift=7.5cm]

\coordinate (Z)   at (0,0);
\coordinate (ka)  at (-3.75,1.6);
\coordinate (kb)  at (-1.25,1.6);
\coordinate (kc)  at (1.25,1.6);
\coordinate (kd)  at (3.75,1.6);
\coordinate (sab) at (-3.75,3.6);
\coordinate (sbc) at (-1.25,3.6);
\coordinate (sca) at (1.25,3.6);
\coordinate (sbd) at (3.75,3.6);


\draw[lit]  (ka) -- (Z);
\draw[lit]  (kb) -- (Z);
\draw[lit]  (kc) -- (Z);
\draw[litU] (kd) -- (Z);

\draw[dom] (sab) -- (ka);
\draw[dom] (sab) -- (kb);
\draw[dom] (sbc) -- (kb);
\draw[dom] (sbc) -- (kc);
\draw[dom] (sca) -- (kc);
\draw[dom] (sca) -- (ka);
\draw[dom] (sbd) -- (kb);
\draw[dom] (sbd) -- (kd);

\draw[prize]  (sab) to[bend left=18]  (Z);
\draw[prize]  (sbc) to[bend right=20] (Z);
\draw[prize]  (sca) to[bend right=18] (Z);
\draw[prizeU] (sbd) to[bend right=18] (Z);

\node[sink]    at (Z)   {$z$};
\node[vertex]  at (ka)  {$t^1_a$};
\node[vertex]  at (kb)  {$t^1_b$};
\node[vertex]  at (kc)  {$t^1_c$};
\node[vertexU] at (kd)  {$t^1_d$};
\node[source]  at (sab) {$s_{\{a,b\}}$};
\node[source]  at (sbc) {$s_{\{b,c\}}$};
\node[source]  at (sca) {$s_{\{c,a\}}$};
\node[source]  at (sbd) {$s_{\{b,d\}}$};

\end{scope}

\end{tikzpicture}

    \caption{
    Example of a reduction from $k$-Clique to \DVP for $k=3$. The green and red edges represent the optimal probes.
    }
    \label{fig:k-clique-reduction}
\end{figure}
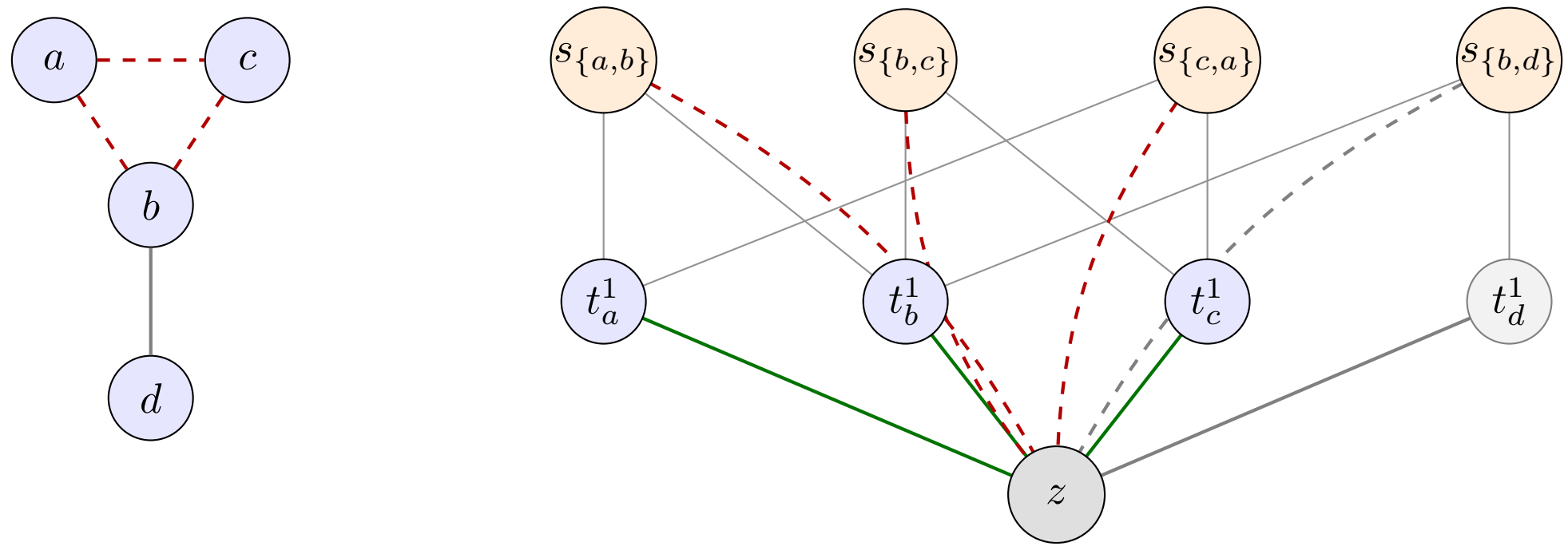

To derive the approximation hardness of \DVP{}, we state the currently sharpest known (parametrized) inapproximability results known for \DkS.

\begin{proposition}\label{prop:hardness-DkS}
    We have the following hardness results for \DkS:
    \begin{enumerate}[label=\roman*)]
        \item\ant{\label{item:hardness-almost-eth} \cite[Theorem~1]{manurangsi2026almost}:
        Assuming the Almost-ETH, for any $\varepsilon \in (0,1)$ there is no polynomial-time algorithm such that, given an integer $k$ and any graph $H=(U,F)$ on $n$ vertices that contains at least one $k$-clique, always output $S \subseteq U$ with $|S|=k$ such that $\density(S) > 2^{-(\log n)^{1-\varepsilon}}$.}
        \item\label{item:hardness-fpt-spch} \cite[Theorem~2]{manurangsi2021strongish}:
        Assuming the SPCH, for every function $f(k) = o(k)$ and every function $g$, there is no $g(k) \cdot n^{O(1)}$-time algorithm such that, given an integer $k$ and any graph \wes{$H=(U,F)$} on $n$ vertices that contains at least one $k$-clique, always outputs $S \subseteq U$ with $|S|=k$ such that $\density(S) \geq \frac{1}{f(k)}$.
    \end{enumerate}
\end{proposition}

\begin{proof}[Proof of Theorem~\ref{thm:dks}]
	Let $S^*\wes{\subseteq U}$ be an optimal solution of the \DkS instance $H = (U,F)$. First, we guess the optimal number of edges $m^* \coloneqq |F[S^*]|$%
    \footnote{Here, \emph{guessing} refers to iterating over all possible values and selecting the result from the iteration with the highest density.}
    and set $D = \lceil m^*/k \rceil$ and $T = kD + m^*$.
	
	Let now $\OPT\wes{\subseteq E_H}$ be the optimal solution of the \DVP instance $G_H\wes{=(V_H,E_H)}$.
	Probing all the edges in the set $\{s_{e}\}_{e\in F[S^*]} \cup \{q^i_u\}_{u\in S^*}^{i\in [D]}$ needs exactly $T$ probes and yields a profit $m^*$, so $w(\OPT) \geq m^*$.

    \paragraph{The reduction.}
	Assume now that there is an $\alpha$-approximation algorithm for \DVP that probes $E' \subseteq E_H$ with $w(E') \geq w(\OPT)/\alpha \geq m^*/\alpha$.
    \ant{For every edge $p_{e,u}^i \in E'$, we remove $p_{e,u}^i$ from $E'$ and add $q_u^i$ to $E'$ (unless it is already in $E'$) \wes{at no additional probing cost}.
    The resulting set $E''$ is feasible since $|E''| \leq |E'|$ and we have $w(E'') = w(E')$.}
	Let $S = \{u\in U:\; q_u^i \in \ant{E''} \ \forall i\in [D]\}$.
	We have $|S| \leq T/D = k + m^*/D \leq 2k$.
	Notice that $s_{\{u,v\}} \in \ant{E''}$ implies $u, v \in S$ and therefore $\{u,v\} \in F[S]$, so $|F[S]| \geq w(\ant{E''}) \geq m^*/\alpha$.
	Now take a subset $S' \subseteq S$ of size $k$ uniformly at random.
    For every $e\in F[S]$, we have $\pr(e\in F[S']) \geq \frac{k-1}{2(2k-1)} = \frac{1}{4} - O\big(\frac{1}{k}\big)$, so
    $\EE\big[|F[S']|\big] \geq \frac{k-1}{2(2k-1)}\cdot\frac{m^*}{\alpha}$ and therefore $(4+O(1/k))\cdot\alpha \cdot \density(S') \geq \density(S^*)$ in expectation.
	The computation of $S' \subseteq S$ with maximum density can be derandomized in time $O(k^2n^{O(1)})$ via the method of conditional expectations, see~\cite{mitzenmacher2017probability}.

    \paragraph{Proof of \ref{item:poly-almost-eth}.}
    \ant{
    Let $\delta \in (0,1)$ and assume that there is a polynomial-time algorithm that outputs, for any \DVP{} instance $G=(V,E)$ (without loss of generality with $w_e\in \{0,1\}$, $p_e=\tau_e=1$ for every $e\in E$), a feasible solution $E' \subseteq E$ with $w(E') \geq w(\OPT) \cdot 2^{-(\log n)^{1-\delta}}$, where $\OPT\subseteq E$ is an optimal solution of the instance.
    }
    
    Consider any \DkS{} instance $H=(U,F)$ with $\ell = |U| > \max\big\{2^{2^{2/\delta}+1},2^{2^{1/(1-\delta)}}\big\}$ which contains a $k$-clique $S^*$, i.e., $\density(S^*) = 1$, taking $k$ large enough such that $\frac{k-1}{2(2k-1)} \geq \frac{1}{5}$.
    We have $n = |F| + D\ell +1 \leq \binom{\ell}{2} + \lceil\frac{\ell-1}{2}\rceil \ell+1 \leq \ell^2$.
    Applying the assumed algorithm on the reduction $G_H$, and using the bounds above, for $\ell$ large enough, we get a set $S' \subseteq U$ with
    \begin{equation*}
        \density(S') \geq  2^{-(\log n)^{1-\delta}} \cdot \frac{1}{5} \density(S^*)
        \geq \frac{1}{5} \cdot 2^{-(\log n)^{1-\delta}}
        \geq \frac{1}{5} \cdot 2^{-(\log n)^{1-\delta}}
        \geq \frac{1}{5} \cdot 2^{-(\log \ell^2)^{1-\delta}}
        > 2^{-(\log \ell)^{1-\delta/2}},
    \end{equation*}
    which is a contradiction to \ref{item:hardness-almost-eth} in Proposition~\ref{prop:hardness-DkS}.
    The last inequality follows from
    \begin{equation*}
        \frac{2^{(\log \ell)^{1-\delta/2}}}{5 \cdot 2^{(\log \ell^2)^{1-\delta}}}
        = \frac{1}{5}\cdot2^{(\log \ell)^{1-\delta/2}-(2\log \ell)^{1-\delta}}
        = \frac{1}{5}\cdot2^{(\log \ell)^{1-\delta}\big((\log \ell)^{\delta/2} -2^{1-\delta}\big)}
        \geq \frac{1}{5} \cdot 2^{2 \cdot (4-2^{1-\delta})} > 1.
    \end{equation*}
    
    \paragraph{Proof of \ref{item:fpt-spch}.}
    Consider any functions $f(n) = o\big(\sqrt{n}\big)$ and $g$, assume without loss of generality that $f$ is monotonic increasing.
    Assume there is an algorithm with running time $g(T) \cdot n^{O(1)}$, that outputs, for any \DVP instance $G=(V,E)$ (without loss of generality with $w_e \in \{0,1\}$, $\tau_e=p_e = 1$ for every $e\in E$) a feasible solution $E' \subseteq E$ with $w(E') \geq w(\OPT)/f(T)$, where $\OPT$ is the optimal solution of the instance $G$.
    
    Consider any \DkS instance $H=(U,F)$ which contains a $k$-clique $S^*$, i.e., $\density(S^*) = 1$, taking $k$ large enough to have $\frac{k-1}{2(2k-1)} \geq \frac{1}{5}$.
    Applying the assumed algorithm on the reduction $G_H$, we get a set $S' \subseteq U$ with
    \begin{equation*}
        \density(S') \geq \frac{1}{5f(T)} \cdot \density(S^*) > \frac{1}{5f(k^2)}
    \end{equation*}
    using the monotonicity of $f$ and the bound $T < k^2$.
    We have $\lim_{n \rightarrow \infty}\frac{5f(n^2)}{n} = \lim_{q\rightarrow\infty} \frac{5f(q)}{\sqrt{q}} = 0$, so assuming the SPCH, we contradict \ref{item:hardness-fpt-spch} in Proposition~\ref{prop:hardness-DkS}.
\end{proof}

\anote{remark $o(\sqrt{T})$ in DVP vs $o(k)$ in \DkS?}

\section{The approximation algorithm}


As assumed in Section~\ref{sec:introduction}, the only condition for a probing policy to be valid is that at any time an edge is probed, it must  lie in the transitive reduction of \ant{the remaining graph and be unprobed}.
We give stronger static criteria to assess whether a subset or a specific edge $e\in E$ can be extracted by a valid policy or not.

\begin{claim}\label{claim:monotonicity-reduction}
    For any acyclic edge set $E$ and any edge $e\in E$, we have $\treduc(E) \setminus \{e\} \subseteq \treduc(E\setminus\{e\})$.
\end{claim}

\begin{proof}
    Let $e\in E$ and $e'=(u,v) \in \treduc(E)\setminus\{e\}$.. Since there is no $u$-$v$-path except $(u,v)$ in $E$, there is  also no $u$-$v$-path except $(u,v)$ in $E\setminus \{e\}$, so $e' \in \treduc(E\setminus\{e\})$.
\end{proof}

We call a set $R = \{e_1, \dots, e_k\} \subseteq E$ \emph{obtainable} if the elements of $R$ can be ordered in a way that $e_1 \in \treduc(E)$ and $e_t \in \treduc(E \setminus \{e_1, \dots, e_{t-1}\})$ for every $t \in \{2, \dots, k\}$.
\ant{By the following lemma, obtainability is the essential static criterion for a set to be probable.}

\begin{lemma}\label{lem:obtainable-policy-criterion}
    For any valid probing policy, the set of successfully probed edges is always obtainable.
    Conversely, for any obtainable set $R \subseteq E$, assuming \ant{$\tau(R)\leq T$ and} $p_e > 0$ for every $e\in R$, there is a valid probing policy that successfully probes every edge in $R$ with non-zero probability.
\end{lemma}

\begin{proof}
    The first part is clear from Claim~\ref{claim:monotonicity-reduction} since any failed probed edge is never removed from $E$.
    Consider now an obtainable set $R = \{e_1, \dots, e_k\} \subseteq E$ with $p_e > 0$ for every $e\in R$ sorted appropriately. Consider the policy $\pi$ that probes the elements of $R$ in the given order and halts as soon as the next candidate is not in the transitive reduction of the remaining graph or once all elements of $R$ have been probed.
    Since $\pi$ only probes edges in $R$ and $\tau(R)\leq T$ by assumption, the budget never runs out before probing all edges in $R$.
    In the event $A= \bigcap_{i=1}^k \{\Omega_{e_i} = 1\}$, $\pi$ probes every element of $R$ successfully.
    Since the random variables $\{\Omega_{e_i}\}_{i\in [k]}$ are independent, we have $\pr(A) = \prod_{i=1}^k p_{e_i} > 0$ and the claim follows.
\end{proof}

\begin{lemma}\label{lem:obtainable-path-criterion}
	For any acyclic edge set $E$ and any set $R \subseteq E$, $R$ is obtainable if and only if there is no $u$-$v$-path in $E\setminus R$ for every edge $(u,v) \in R$.
\end{lemma}

\begin{proof}
	We argue by induction in $|R|$.
	The case $|R|=1$ is precisely the definition of $\treduc(E)$.
	Assume now $|R| > 1$.
	If $R$ is obtainable and $e=(u,v)$ is the last element that was added to $R$, $R' = R \setminus \{e\}$ is obtainable by assumption and there is no $u'$-$v'$-path in $E\setminus R$ for any $(u',v') \in R'$ by induction.
    Since $e\in \treduc(E\setminus R')$, there is also no $u$-$v$-path in $E\setminus R'$, so also none in $E\setminus R = E\setminus (R' \cup \{e\})$.
	
	Assume now that there is no $u$-$v$-path in $E\setminus R$ for every $e=(u,v) \in R$.
	Let $e'=(u',v') \in R$ be the edge in $R$ such that the $u'$-$v'$-path in $E$ is the longest among all edges in $R$. 
	Let $R' = R \setminus \{e'\}$.
	For any $(u,v)\in R'$, $e'$ cannot be in any $u$-$v$-path since we could construct a $u$-$v$-path longer than the $u'$-$v'$-path, so there is no $u$-$v$-path in $E\setminus R'$, meaning that $R'$ is obtainable by induction. 
    By the same argument, $e' \in \treduc(E \setminus R')$, so $R$ is obtainable.
\end{proof}

With this criterion, we can draw a similarity between obtainability and the Min-Cut problem.
We recall that for any directed graph $G=(V,E)$ with $\{u,v\} \subseteq V$, an edge set $C \subseteq E$ is an \emph{$u$-$v$-cut} if there is a vertex set $\{u\} \subseteq U \subseteq V\setminus\{v\}$ such that $C = \delta^+(U)$, where $\delta^+(U)$ denotes the set of all edges from $U$ to $V\setminus U$.

\begin{proposition}\label{prop:FUS-cut}
    Let $E$ be an acyclic edge set and let $e=(u,v) \in E$.
    Any obtainable set $R \subseteq E$ with $e\in R$ contains a $u$-$v$-cut and any $u$-$v$-cut is obtainable.
\end{proposition}

\begin{proof}
    Let $R \subseteq E$ with $e\in R$ be obtainable. By Lemma~\ref{lem:obtainable-path-criterion}, there are no $u$-$v$-paths in $E\setminus R$.
    Let $U = \{w\in V:\; \exists\text{ $u$-$w$-path in $E \setminus R$}\}$.
    Clearly, $\delta^+(U)$ is a $u$-$v$-cut since $u \in U$ and $v\notin U$ and $\delta^+(U) \subseteq R$.

    Let $C$ be an $u$-$v$-cut with $\{u\} \subseteq U \subseteq V\setminus \{v\}$ such that $C = \delta^+(U)$.
    Assume that $C$ is not obtainable, meaning by Lemma~\ref{lem:obtainable-path-criterion} there is a\wes{n edge} $e' = (u',v') \in C$ and an $u'$-$v'$-path $P\subseteq V$ with no edge in $C$.
    Since $e' \in \delta^+(U)$, we have $u'\in U$ but $v' \notin U$.
    Let $w \in P$ be the last vertex in $P$ such that the subpath of $P$ from $u'$ to $w$ lies entirely in $U$.
    The vertex $w$ exists since $u' \in U$ and $w\ne v'$ since $v'\notin U$.
    Therefore the successor $w' \in P$ of $w$ in $P$ exists and $w' \notin U$, in particular $(w, w') \in \delta^+(U) = C$, which is a contradiction.
\end{proof}

It follows immediately from Proposition~\ref{prop:FUS-cut} that an inclusion-minimal obtainable set containing $e = (u,v)$ is an inclusion-minimal $u$-$v$-cut and vice versa.
In particular, computing an optimal minimum-cost obtainable set including $e$ is equivalent to computing a minimum $u$-$v$-cut on the same graph with capacities $(\tau_e)_{e\in E}$, which is achievable in time $O(nm)$, see~\cite{orlin2013max}.

For the analysis of a policy, we first need to assure that some probabilistic notions are well-defined, especially for an optimal policy.
While we allow policies to sample other random variables than $\{\Omega_e\}_{e\in E}$, we ensure that this is not necessary for optimality.
\wnote{I don't really understand this. With 'other variables', is it non-bernoulli? if so, are there more than success/failure outcomes?}
For any policy $\pi$ and any $e\in E$, let $A_{e,\pi}$ refer to the event that $\pi$ probes $e$ and let $B_e \coloneqq \{\Omega_e=1\}$.

\begin{lemma}\label{lem:opt-dvp-events}
    For every \DVP instance $G=(V,E)$, there exists an optimal policy $\pi^*$ that only depends on the variables $\{\Omega_e\}_{e\in E}$ and such that the events $A_{e,\pi^*}$ and $B_e$ are independent for every $e\in E$.
\end{lemma}
\begin{proof}
    For a \DVP instance $G=(V,E)$, let $[H,F,T]$ denote a stage of a probing process where $H$ is the set of unprobed edges, $F$ of failed probed edges and $T$ the leftover budget.
    Let $\calx(H,F,T) \coloneqq \treduc(H\cup F) \cap \{e\in H:\; \tau_e\leq T\}$ be the set of probable edges and let $W(H,F,T)$ be the optimal expected profit.
    As argued in \cite{finnah2025overcoming}, $W(H,F,T)$ is computable with the Bellman recursion $W(H,F,T)=0$ if $\calx(H,F,T) = \emptyset$ and
    \begin{equation}\label{eq:bellman-dvp}
        W(H,F,T ) = \max_{e\in \calx(H,F,T)} \Big( p_e\big(w_e+ W(H\setminus\{e\},F,T-\tau_e)\big) + (1-p_e)W(H\setminus\{e\},F\cup\{e\},T-\tau_e) \Big)
    \end{equation}
    otherwise.
    Therefore, we can select $\pi^*$ to be a policy that, at every stage $[H,F,T]$, picks an edge that maximizes \eqref{eq:bellman-dvp}, breaking ties deterministically, meaning that $\pi^*$ depends only on $\{\Omega_e\}_{e\in E}$.
    
    For a stage $S=[H,F,T]$, let $C_S$ be the event that the policy attains the stage $S$. 
    By induction in the number of probes, for every stage $S=[H,F,T]$, we have either $C_S=\emptyset$ or $C_S=\bigcap_{e\in E\setminus(H\cup F)}B_e \cap \bigcap_{e'\in F} \overline{B_{e'}}$.
    \wnote{(NEW nov 3rd): $e'\in F$ used as variable, but not used. \ant{fixed}}
    In both cases, for every $e\in H$, $B_e$ and $C_S$ are independent. If $S$ and $S'$ are two stages where $\pi^*$ picks $e$, we have $C_S \cap C_{S'} = \emptyset$ since an edge is never picked twice.
    \wnote{(NEW nov 3rd): Should it be $C_S\cap C_{S'}=\emptyset$? Otherwise neither event ever is attained right? \ant{fixed}}
    Since the event $A_{e,\pi^*}$ is the event that such a stage is attained, $A_{e,\pi^*}$ and $B_e$ are independent.
\end{proof}

It follows that the expected profit of any policy $\pi$ can be formally reformulated as $\sum_{e\in E}w_e\pr(A_{e,\pi} \cap B_e) = \sum_{e\in E}w_e\pr(A_{e,\pi})\pr(B_e)$.

\paragraph{An adaptive policy for uniform probabilities.}
If a \DVP instance has exactly one edge $e$ with $w_e>0$, an optimal policy for that instance is coincidentally a policy which has the highest chance of probing $e$.
By Lemma~\ref{lem:opt-dvp-events}, we can assume that this policy only depends on $\{\Omega_e\}_{e\in E}$.
For any policy $\pi$, we write $\theta(e,\pi) \coloneqq \pr(A_{e,\pi})$ and call $\pi$ \emph{$e$-optimal} if $\pi \in \argmax_{\pi'}\theta(e,\pi')$, i.e., if $\pi$ has the highest chance of probing $e$ among all feasible policies.

Intuitively, for any (globally) optimal policy $\pi^*$, if we know the edge $e^*$ with highest expected profit among all edges in $E$ and we know an $e^*$-optimal policy $\pi$, $\pi$ will extract at least as much (expected) profit from $e^*$ as $\pi^*$ but potentially nothing from the other $m-1$ edges; therefore, uniformly sampling a single edge $e \in E$ and running an $e$-optimal policy is an $m$-approximation (in expectation) of an optimal policy.

We now derive a simple $e'$-optimal policy for any edge $e'$ in a DVP instance $G=(V,E)$ in the case of uniform costs and probabilities.

\begin{lemma}\label{lem:e-optimal-policy}
    Let $p\in (0,1]$ and let $G=(V,E)$ be a \DVP{} instance with $\tau_e = 1$ and $p_e = p$ for every $e\in E$.
    Consider an edge $e = (s,t) \in E$.
    A policy which, at any step, probes an edge on a minimum $s$-$t$-cut (consisting of unprobed edges only), is $e$-optimal, regardless of the choice of the cut and the edge.
\end{lemma}

\begin{proof}
    Without loss of generality, we assume $T < m$ since otherwise, probing any edges (as long as any edge is probable) is trivially an optimal policy. 
    
    We again denote a stage of the probing process where $H$ is the set of unprobed edges, $F$ the set of failed probed edges and $T$ the leftover budget as $[H,F,T]$, or $[H,F]$ if we omit the budget.
    If we probe $e \in H$ at the stage $[H,F,T]$, if the probe succeeds, we are at stage $[H\setminus\{e\},F,T-1]$, and at $[H\setminus\{e\},F\cup\{e\},T-1]$ otherwise.
    At stage $[H,F]$, let $\sigma(H,F)$ be the size of a minimum admissible $s$-$t$-cut. \ant{We say that an $s$-$t$-cut is \emph{admissible} if it which contains only edges in $H$;}
    We set $\sigma(H,F) = +\infty$ if no admissible $s$-$t$-cut exists, i.e., if $F$ contains an $s$-$t$-path.%
    \footnote{A minimum admissible $s$-$t$-cut can be found on the graph $(V,H\cup F)$ by assigning weights $\tau'_e= 1$ for every $e \in H$ and $\tau'_e = m+1$ for every $e\in F$ and computing a minimum $s$-$t$-cut $C$. $C$ is admissible if and only if $\tau'(C) \leq m$; otherwise, we have $\sigma(H,F) = +\infty$.}

    At any stage $[H,F]$, let $e\in H$ be in a minimum admissible $s$-$t$-cut $C$. $C \setminus \{e\}$ is an admissible $s$-$t$-cut at $[H\setminus\{e\},F]$, so $\sigma(H\setminus\{e\},F) \leq \sigma(H,F)-1$. However, if $C'$ is an admissible $s$-$t$-cut at $[H\setminus\{e\},F]$, $C'\cup \{e\}$ is a cut at $[H,F]$, so $\sigma(H\setminus\{e\},F) = \sigma(H,F)-1$.
    We have $\sigma(H\setminus\{e\},F\cup\{e\}) \geq \sigma(H,F)$ since any admissible $s$-$t$-cut at $[H\setminus\{e\},F\cup\{e\}]$ is an admissible $s$-$t$-cut at $[H,F]$.
    
    At $[H,F]$, let $e\in H$ not lie on any admissible minimum $s$-$t$-cut.
    Any minimum admissible $s$-$t$-cut $C$ (without $e$) at $[H,F]$ is also a minimum admissible $s$-$t$-cut at $[H\setminus\{e\},F\cup\{e\}]$ and vice versa, so $\sigma(H\setminus\{e\},F\cup\{e\}) = \sigma(H,F)$.
    Let $C'$ be a minimum admissible $s$-$t$-cut at the stage $[H\setminus\{e\},F]$.
    Since $C'$ disconnects $s$ and $t$ in $(H \setminus\{e\})\cup F$, $C'\cup\{e\}$ disconnects $s$ and $t$ in $H\cup F$ as well. However, since $e$ lies on no minimum admissible $s$-$t$-cut at the stage $[H,F]$, $C'\cup\{e\}$ has no minimum cardinality at stage $[H,F]$, so we have $\sigma(H\setminus\{e\},F) = |C'| > \sigma(H,F)-1$. Since any admissible $s$-$t$-cut (without $e$) at $[H, F]$ is an admissible $s$-$t$-cut in $[H\setminus \{e\}), F]$, we have $\sigma(H\setminus\{e\},F) \leq \sigma(H,F)$, so we get $\sigma(H\setminus\{e\},F) = \sigma(H,F)$.
    
    Let $P^*(H,F,T)$ be the success probability of an optimal policy at that stage.
    By Proposition~\ref{prop:FUS-cut}, the algorithms succeed if we managed to disconnect $s$ and $t$, so we have $P^*(H,F,T)=1$ if and only if $\sigma(H,F)=0$.
    As shown above, $\sigma(H,F)$ can only decrease by $1$ at each probe, we have $P^*(H,F,T)=0$ if $T < \sigma(H,F)$.
    For any $e\in H$, let $P_e(H,F,T)$ be the success probability of an optimal policy at the next stage, assuming $e$ is picked at the current stage.
    Therefore, $P_e(H,F,T) = p \cdot P^*(H\setminus \{e\},F,T-1) + (1-p)\cdot P^*(H\setminus\{e\},F\cup\{e\},T-1)$.
    In particular, a policy is optimal if, at any stage $(H,F,T)$, it probes an arbitrary edge in $\argmax_{e\in H} P_e(H,F,T)$.%
    
    By induction in $T$, we show that at any stage $[H,F,T]$ with $T \geq \sigma(H,F)>0$, any edge on any minimum admissible $s$-$t$-cut is in $\argmax_{e\in H} P_e(H,F,T)$; the lemma follows immediately.
    If $T=1$, we have $\sigma(H,F)=1$, and the only way to reduce $\sigma(H,F)$ is to pick the edge on the (unique) admissible single-edge $s$-$t$-cut.
    Assume now $T>1$.
    
    At a stage $[H,F,T]$, let $f\in \argmax_{e\in H} P_e(H,F,T)$ be the edge chosen by an optimal policy.
    If $f$ belongs to a minimum admissible $s$-$t$-cut, the claim follows directly by induction; otherwise, let $C$ be any minimum admissible $s$-$t$-cut. Since $C$ is obtainable by Proposition~\ref{prop:FUS-cut}, there exists some edge $e'\in \treduc(H \cup F) \cap C$.
    As argued above, $C$ is a minimum admissible $s$-$t$-cut at the stages $[H\setminus \{f\}), F]$ and $[H\setminus \{f\}), F\cup\{f\}]$, so $e' \in \argmax_{e\in H} P_e(H\setminus \{f\}, F,T-1)$ and $e' \in \argmax_{e\in H} P_e(H\setminus \{f\}, F\cup\{f\},T-1)$ by induction.
    We get
    \begin{align*}
        P^*(H,F,T)
        &= P_f(H,F,T) \\
        &= p \cdot P^*(H\setminus \{f\},F,T-1) + (1-p)\cdot P^*(H\setminus\{f\},F\cup\{f\},T-1) \\
        &= p \cdot P_{e'}(H\setminus \{f\},F,T-1) + (1-p)\cdot P_{e'}(H\setminus\{f\},F\cup\{f\},T-1) \\
        &= p^2 \cdot P^*(H\setminus \{e',f\},F,T-2) \\
        &\quad + p(1-p)\cdot \big(P^*(H\setminus\{e',f\},F\cup\{f\},T-2) + P^*(H\setminus\{e',f\},F\cup\{e'\},T-2)\big)\\
        &\quad+ (1-p)^2 \cdot P^*(H\setminus\{e',f\},F\cup\{e',f\},T-2) \\
        &= p \cdot P_f(H\setminus \{e'\},F,T-1) + (1-p)\cdot P_f(H\setminus\{e'\},F\cup\{e'\},T-1) \\
        &\leq p \cdot P^*(H\setminus \{e'\},F,T-1) + (1-p)\cdot P^*(H\setminus\{e'\},F\cup\{e'\},T-1) \\
        &= P_{e'}(H,F,T),
    \end{align*}
    meaning $e'\in \argmax_{e\in H} P_e(H,F,T)$.
    Since the choice of $C$ and $e'$ was arbitrary, the proof is complete.
\end{proof}

\begin{proof}[Proof of Theorem~\ref{thm:adaptive}]
    Let $Z$ be a random variable which is uniformly distributed over $E$ and independent of $\{\Omega_e\}_{e\in E}$.
    We consider the policy $\tilde \pi$ which first samples an edge $\tilde e$ from $Z$, then runs the $\tilde e$-optimal policy $\pi_{\tilde e}$ of Lemma~\ref{lem:e-optimal-policy}.
    Since $A_{e,\tilde \pi} \supseteq\{Z=e\} \cap A_{e,\pi_e}$, and the events $\{Z= e\}$, $A_{e, \pi_e}$ and $B_{e}$ are independent for every $e\in E$, we get expected profit
    \begin{align*}
        W(\tilde \pi) &=
        \sum_{e\in E}w_e\pr(B_e \cap A_{e,\tilde \pi}) \\
        &\geq\sum_{e\in E} w_e \pr(Z=e)\pr(B_e) \pr(A_{e, \pi_e})
        = \sum_{e\in E} \frac{pw_{e}}{m} \theta(e, \pi_e)
        \geq \frac{1}{m} \sum_{e\in E} pw_{e} \theta(e,\pi^*)
        = \frac{\opt}{m}.
    \end{align*}
    Any sampled $e$-optimal policy makes up to $T$ probes (since we assumed $T<m$) and updates the transitive reduction and the minimum cut at each iteration in time $O(m+n^{1.528})$ and $O(mn)$ respectively, yielding a total time of at most $O(Tmn)$.
\end{proof}

\paragraph{A policy for general instances.}
The first step of the algorithm is, for every $e\in E$, to compute a minimum-cost obtainable set $S_e \subseteq E$ with $e\in S_e$.
Recall that we also assume $p_e > 0$ for every $e\in E$ since we can assign cost $\tau_e = T+1$ to any edge with $p_e=0$ since an optimal policy would never probe an edge with success probability $0$.

Every edge $e\in E$ is treated as an object with size $\tilde \tau_e \coloneqq \tau(S_e)$ and profit $\tilde w_e \coloneqq w_e\prod_{f\in S_e}p_f$, and the goal is to pick objects with maximum profit but bounded cumulative size, like a Knapsack instance.
We assign to every $e\in E$ a \emph{cost efficiency ratio $r_e \coloneqq \frac{\tilde w_e}{\tilde \tau_e}$}.

We order the edges in $E$ in decreasing cost efficiency ratio \ant{but exclude the edges with $\tilde \tau_e > T$}, thus having $E = \{e_1, e_2,  \dots, e_m\}$ with $r_{e_k} \geq r_{e_{k+1}}$ for every $k \in [m-1]$.
Let $\kappa = m$ if $\sum_{i=1}^{m} \tilde\tau_{e_i} \leq T$ and otherwise be the integer with $\sum_{i=1}^\kappa \tilde\tau_{e_i} \leq T < \sum_{i=1}^{\kappa+1} \tilde\tau_{e_i}$.
Let also $\bar e$ be the edge with maximum profit, i.e., $\bar e \in \argmax_{e\in E}\tilde w_e$.

From this point on, the algorithm is greedy.
We set $E' \coloneqq \bigcup_{i=1}^\kappa S_{e_i}$ if $\sum_{i=1}^{\kappa} \tilde\tau_{e_i} \geq \tilde\tau_{\bar e}$ and $E' \coloneqq S_{\bar e}$ otherwise.
Then, we probe any probable edge in $E'$ in arbitrary order until no edge in $E'$ can be probed, i.e., all unprobed edges in $E'$ are not in the transitive reduction of the resulting graph.

\begin{proof}[Proof of Theorem~\ref{thm:greedy}]
    Let $\pi$ denote the policy described above.
    In the case $\sum_{i=1}^{m} \tilde\tau_{e_i} \geq \tau_{\tilde e}$, notice that for any $i\in [\kappa]$, in the event $\bigcap_{f\in S_{e_i}\setminus\{e\}}B_f$, $e_i$ is successfully probed; in particular, we have $\theta(e_i,\pi) \geq \pr\big(\bigcap_{f\in S_{e_i}\setminus\{e_i\}}B_f\big) = \frac{1}{p_{e_i}}\prod_{f\in S_{e_i}} p_f$.
    By the same argument, we have $\theta(\bar e,\pi) \geq \frac{1}{p_{\bar e}}\prod_{f\in S_{\bar e}} p_f$ when $\sum_{i=1}^{m} \tilde\tau_{e_i} < \tau_{\tilde e}$.
    
    Let $\pi^*$ be an optimal (adaptive) probing policy as in Lemma~\ref{lem:opt-dvp-events} and let $x^* = \theta(e,\pi^*)$ for every $e\in E$.
    We have $W(\pi^*)= \sum_{e \in E}p_e w_e x_e^*$.
    Since $\pi^*$ is feasible, the cumulative cost of the edges probed by $\pi^*$ is never above $T$, i.e., $\sum_{e\in E} \tau_e \indic_{A_e} \leq T$.
    This must thus hold in expectation as well, so $T \geq \EE\big[\sum_{e\in E} \tau_e \indic_{A_e} \big] = \sum_{e \in E} \tau_e x_e^*$.
    Consider the linear program
    \begin{equation}\label{eq:greedy-lp}
        L \coloneqq \max \Big\{\sum_{e\in E} \tilde w_e  y_e\; :\
        0 \leq y \leq 1,\
        \sum_{e\in E} \tilde\tau_e y_e \leq T \Big\}
    \end{equation}
    For every $e\in E$, notice that $(x^*_e/T)_{e \in E}$ is a feasible fractional solution with respect to the LP \eqref{eq:greedy-lp} since $\sum_{e\in E}\tau(S_e) (x^*_e/T) \leq \sum_{e\in E} x^* \leq \sum_{e\in E} \tau_e x^*_e \leq T$.
    Also, $|S_e \setminus\{e\}| \leq \tau(S_e) \leq T$, so $\prod_{f\in S_e\setminus\{e\}}p_f \geq p_{\min}^{|S_e|-1} \geq p_{\min}^{T-1}$ and $\tilde w_e \geq p_{\min}^T w_e$.
    Together, we get $L \geq \frac{p_{\min}^T}{T}W(\pi^*)$.

    One fractional optimum $y^*$ of \eqref{eq:greedy-lp} is $y^*_{e_i} = 1$ for $i\leq \kappa$, $y^*_{e_{\kappa+1}} = 1 -\sum_{i=1}^\kappa y^*_{e_{i}}$ and $y^*_{e_i} = 0$ for $i > \kappa+1$, with $E$ sorted as above \ant{since \eqref{eq:greedy-lp} has the form of the canonical LP relaxation of the 0-1 Knapsack problem} \cite{martello1990knapsack}.
    In particular, we have $\sum_{i=1}^\kappa \tilde w_{e_i} + \tilde w_{\bar e} \geq \sum_{i=1}^{\kappa+1} \tilde w_{e_i}\geq L$.

    Assume now that $\sum_{i=1}^\kappa \tilde w_{e_i} \geq \tilde w_{\bar e}$, the other case is symmetric.
    By construction, for any $i \in [\kappa]$, $e_i$ is probed if $\prod_{f\in S_{e_i}} \Omega_{f} = 1$, so $e_i$ is probed and succeeds with probability at most $p_{e_i}\prod_{f\in S_{e_i}}p_f$.
    By the linearity of expectation, the total expected profit is
    \begin{equation*}
        \ant{W(\pi) =}
        \sum_{i=1}^\kappa w_{e_i} p_{e_i}\theta(e_i,\pi)
        = \sum_{i=1}^\kappa \tilde w_{e_i}
        \geq \frac{1}{2} \Big( \sum_{i=1}^\kappa \tilde w_{e_i} + \tilde w_{\bar e} \Big)
        \geq \frac{L}{2}
        \geq \frac{p_{\min}^T}{2T}W(\pi^*).
    \end{equation*}
    \ant{
    Computing $S_e$ for every $e\in E$ can be achieved in time $O(m^2n)$ since it consists of computing a Min-Cut instance for every edge.
    Then, during the probing process, $\treduc(E)$ can be updated in time $O(m+n^{1.528})$ at each probe, and the choice of the probe in $E'$ is arbitrary, the total running time is bounded by $O(m^2n)$.
    }
\end{proof}

\section{The algorithm for total orders}

We give an FPT algorithm parameterized in the budget $T$ for instances $G=(V,E)$ that form a tournament, as defined in Section~\ref{sec:introduction}.
For simplicity, we order $V=\{v_1, \dots, v_n\}$ such that for any $i,j\in [m]$, we have $(v_i,v_j)\in E$ if and only if $i < j$.
In this setting, the following lemma follows directly from Lemma~\ref{lem:obtainable-path-criterion}.

\begin{lemma}\label{lem:path_max_span}
    Let $R \subseteq E$ be obtainable.
    For every $e=(v_i,v_j) \in R$, we have $j-i \leq |R|$.
\end{lemma}
\begin{proof}
    By Lemma~\ref{lem:obtainable-path-criterion}, for $(v_i,v_j) \in R$, for any $i<k<j$, the $v_i$-$v_j$-path $(v_i,v_k,v_j)$ cannot be in $E\setminus R$, so $(v_i,v_k)\in R$ or $(v_k,v_j)\in R$.
    In particular, $|R\setminus\{e\}| = j-i-1$, and the claim follows.
\end{proof}

\ant{
The main observation is that for any obtainable set $R \subseteq E$ with $\tau(R) \leq T$, we can divide $G$ into up to $T$ disjoint blocks of $T^2+1$ consecutive vertices such that every edge in $R$ lies in \wes{exactly} one of those blocks.
We define the $(s,t)$-block $B_{s,t}$ to be the edge set of the induced subgraph of $G$ on the vertex set $\{v_{s}, \dots, v_t\}$.
}

\begin{lemma}\label{lem:path-block-decomposition}
    Let $R \in E$ be obtainable with $\tau(R)\leq T$.
    There is a family $\calb$ of up to $T$ disjoint blocks such that $t-s \leq T^2$ for each block $B_{s,t} \in \calb$ and $R \subseteq \bigcup_{B\in \calb}B$.
    Furthermore, for every block $B \in \calb$, $R\cap B$ is obtainable.
\end{lemma}

\begin{proof}
    Consider the graph $H=(\ant{R,Z})$ with $Z$ defined as follows: For every $e=(v_i,v_j),f=(v_k,v_\ell) \in R$, we have $\{e,f\} \in Z$ whenever $j\geq k$ and $\ell\geq i$.

    Let $X=\{(v_{i_1},v_{j_1}), \dots, (v_{i_r},v_{j_r})\} \subseteq R$ be a connected component of $H$ of size $r$, and let $s(X) \coloneqq \min_{t\in[r]}i_t$ and $t(X) \coloneqq \max_{t\in[r]}j_t$.
    By Lemma~\ref{lem:path_max_span}, we have $j_t-i_t \leq |R| \leq T$ for every $t\in [r]$, so $t(X) - s(X) \leq Tr \leq T^2$ \ant{by induction}.
    For $(v_k,v_\ell) \in R\setminus X$, we have either $k> t(X)$ or $\ell < s(X)$.
    In particular, if $Y$ is another connected component of $H$, we have $s(Y) > t(X)$ or $t(Y) < s(X)$, so the blocks $B_{s(X),t(X)}$ and $B_{s(Y),t(Y)}$ are disjoint.
    for any $(v_i,v_j) \in R$ in the connected component $X$ of $H$, any $i$-$j$-path on $E$ lies in $B_{s(X),t(X)}$ meaning that $E\setminus R$ contains an $i$-$j$-path if and only if $B_{s(X),t(X)}\setminus R$ contains an $i$-$j$-path.
    By Lemma~\ref{lem:obtainable-path-criterion}, this means that every connected component of $H$ is obtainable.
    Trivially $H$ has up to $|R| \leq T$ connected components.

    Therefore, if $\cala$ is the set of connected components of $H$, we can choose $\calb = \{B_{s(X),t(X)}\}_{X\in\cala}$ as the family of blocks which satisfies the properties of the claim.
\end{proof}

We now describe the algorithm of Theorem~\ref{thm:path-case}.

\begin{proof}[Proof of Theorem~\ref{thm:path-case}]

    Let $\OPT_{s,t,T'} \in \argmax\{R\subseteq B_{s,t};\; R \text{ is obtainable and } \tau(R)\leq T'\}$ refer to an optimal solution of the \DVP{} instance restricted on the block $B_{s,t}$ and budget $T'$.
    Clearly, $\OPT_{1,n,T}$ is an optimal solution for the entire instance.
    If $t-s \leq T^2$, there are at most $2^{O(T'\log T)}$ many candidates for $\OPT_{s,t,T'}$ since $B_{s,t}$ has $\ell =\binom{t-s+1}{2} \leq \binom{T^2+1}{2} \leq T^4$ edges and there are there are at most $\binom{\ell}{T'} \leq \ell^{T'} \leq T^{4T'} \leq 2^{O(T'\log T)}$ obtainable sets in $B_{s,t}$.
    
    We show now that for every $k\in [n]$ and $T'\in [T]$, $\OPT_{1,k,T'}$ can be selected via the recursion
    \begin{equation}\label{eq:recursion-path-opt}
        \OPT_{1,k,T'} \in \argmax_{k-T^2 \leq s \leq k, T'' \leq T'} \{w(\OPT_{1,s-1,T'-T''} \cup \OPT_{s,k,T''})\}.
    \end{equation}

    First, notice that $\OPT_{1,s-1,T'-T''} \cup \OPT_{s,k,T''}$ is always obtainable by Lemma~\ref{lem:path-block-decomposition}.
    Trivially, we have $\OPT_{1,1,T'} = \emptyset$ since the block $B_{1,1}$ is empty.
    For $k>1$, if there is no edge $(v_i,v_k)$ in $\OPT_{1,k,T'}$, we have $w(\OPT_{1,k,T'}) = w(\OPT_{1,k-1,T'}) = w(\OPT_{1,k-1,T'} \cup \OPT_{k,k,0})$.
    Otherwise, there is a family of boxes $\calb$ with the properties in Lemma~\ref{lem:path-block-decomposition}.
    In particular, since there is an edge $(v_i,v_k)$ in $\OPT_{1,k,T'}$, there is one block $B_{s,k}$ in $\calb$ with $s \geq k-T^2$.
    Let $T'' = \tau(\OPT_{1,k,T'}\cap B_{s,k})$.
    By induction and maximality of $\OPT_{1,k,T'}$,
    \begin{align*}
        w(\OPT_{1,k,T'})
        &= w(\OPT_{1,k,T'}\cap \ant{B_{s,k}}) +  w(\OPT_{1,k,T'}\cap \ant{B_{1,s-1}}) \\
        &\leq w(\OPT_{s,k,T''}) + w(\OPT_{1,s-1,T'-T''}) \\
        &\leq w(\OPT_{1,k,T'}),
    \end{align*}
    so all the inequalities are tight and \eqref{eq:recursion-path-opt} is correct.
    For every $k\in [n]$, $T' \in [T]$, there are at most $T'\cdot T^2 \leq T'$ choices of the last block $B_{s,k}$, and $\OPT_{s,k,T''}$ can be found by enumeration in time $2^{O(T'\log T)}n^{O(1)}$.
    Since we need to compute $nT$ values of $F$, we end up with a total running time $2^{O(T'\log T)}n^{O(1)}$.

\end{proof}

\bibliographystyle{alpha}
\bibliography{dvp-bib.bib}

\end{document}